\documentclass[11pt]{article}
\usepackage{axodraw}
\usepackage{epsfig}
\usepackage{amsfonts}
\usepackage{amsmath}
\usepackage{verbatim}
\usepackage{color}
\usepackage{hyperref}
\allowdisplaybreaks[1]

\input pix.sty

\newcommand{\mW}{m_\rmii{$W$}}
\newcommand{\mZ}{m_\rmii{$Z$}}
\newcommand{\mA}{m_\rmii{$A$}}
\newcommand{\mQ}{m_\rmii{$Q$}}
\newcommand{\mU}{m_\rmii{$U$}}
\newcommand{\tmU}{\tilde{m}_\rmii{$U$}}
\newcommand{\Hc}{{\rm H.c.\ }}

\renewcommand{\eq}{eq.~}
\renewcommand{\eqs}{eqs.~}
\renewcommand{\se}{sec.~}

\renewcommand{\fig}{fig.~}
\renewcommand{\figs}{figs.~}

\newcommand{\Tc}{T_{\rm c}}

\newcommand{\rmO}{{\mathcal{O}}}
\newcommand{\bmu}{\bar\mu}

\def\lsi{\raise0.3ex\hbox{$<$\kern-0.75em\raise-1.1ex\hbox{$\sim$}}}
\def\gsi{\raise0.3ex\hbox{$>$\kern-0.75em\raise-1.1ex\hbox{$\sim$}}}
\newcommand{\lsim}{\mathop{\lsi}}
\newcommand{\gsim}{\mathop{\gsi}}

\newcommand{\rmii}[1]{{\mbox{\tiny\rm{#1}}}}

\newcommand{\Tint}[1]{{\hbox{$\sum$}\!\!\!\!\!\!\!\int\,}_{\!\!\!\!\raise-0.9ex\hbox{$\scriptstyle{#1}$}}}
\newcommand{\Tinti}[1]{{{\Sigma}\!\!\!\!\raise0.3ex\hbox{$\int$}_\rmii{${#1}$}}}

\newcommand{\bi}{\begin{itemize}}
\newcommand{\ei}{\end{itemize}}

\newcommand{\hide}[1]{ }

\newcommand{\deltabar}{\raise-0.02em\hbox{$\bar{}$}\hspace*{-0.8mm}{\delta}}
\makeatletter \@addtoreset{equation}{section} \makeatother

\makeatletter
\renewcommand\section{\@startsection {section}{1}{\z@}%
                                   {-5.5ex \@plus -1ex \@minus -.2ex}
                                   {2.3ex \@plus.2ex}%
                                   {\normalfont\large\bfseries}}
\renewcommand\subsection{\@startsection{subsection}{2}{\z@}%
                                     {-3.25ex\@plus -1ex \@minus -.2ex}%
                                     {1.5ex \@plus .2ex}%
                                     {\normalfont\normalsize\bfseries}}
\renewcommand\thesection {\@arabic\c@section}
\renewcommand\thesubsection   {\thesection.\@arabic\c@subsection}
\renewcommand{\@seccntformat}[1]{%
\csname the#1\endcsname.\hspace{1.0em}}
\makeatother

\begin{document}

\flushbottom

\begin{titlepage}

\begin{flushright}
BI-TP 2012/49\\
HIP-2012-28/TH\\
\vspace*{1cm}
\end{flushright}
\begin{centering}
\vfill

{\Large{\bf
 Lattice study of an 
 electroweak phase transition \\[2mm]
 at $m_h \simeq 126$~GeV }}

\vspace{0.8cm}

M.~Laine$^{\rm a}$, 
G.~Nardini$^{\rm b}$, 
K.~Rummukainen$^{\rm c}$

\vspace{0.8cm}

$^\rmi{a}$%
{\em
Institute for Theoretical Physics, 
Albert Einstein Center, University of Bern, \\ 
Sidlerstrasse 5, CH-3012 Bern, Switzerland\\}

\vspace*{0.3cm}

$^\rmi{b}$%
{\em
Faculty of Physics, University of Bielefeld, 
D-33501 Bielefeld, Germany\\}

\vspace*{0.3cm}

$^\rmi{c}$%
{\em
Department of Physics and Helsinki Institute of Physics, \\
University of Helsinki, P.O.Box 64,  
FIN-00014 Helsinki, Finland\\}

\vspace*{0.8cm}

\mbox{\bf Abstract}
 
\end{centering}

\vspace*{0.3cm}
 
\noindent
We carry out lattice simulations of a cosmological electroweak phase
transition for a Higgs mass $m_h \simeq$~126 GeV. The analysis is
based on a dimensionally reduced effective theory for an MSSM-like
scenario including a relatively light coloured SU(2)-singlet scalar,
referred to as a right-handed stop.  The non-perturbative transition
is stronger than in 2-loop perturbation theory, and may offer a window
for electroweak baryogenesis.  The main remaining uncertainties concern 
the physical value of the right-handed stop mass which according to 
our analysis could be as high as $m^{ }_{\tilde t_R}\simeq$ 155~GeV; 
a more precise effective theory derivation and vacuum renormalization 
than available at present are needed for confirming this value.

\vfill

%
 
\vspace*{1cm}
  
\noindent
November 2012

\vfill

\end{titlepage}

%
\section{Introduction}

One of the most important cosmological conundra is that we live in a
Baryon Asymmetric Universe (BAU), meaning that very few antiprotons are 
seen in cosmic rays, although quarks and antiquarks are produced in almost
equal measure at collider experiments.  One of the possible
explanations for BAU is called electroweak 
baryogenesis~\cite{krs}: it makes use of the Higgs mechanism 
which is assumed to undergo a ``first order'' phase 
transition in the Early Universe. A phase transition 
of this type leads to thermal non-equilibrium, 
one of the necessary Sakharov conditions for explaining
the BAU. The others (C, CP, and baryon number violation) are also
part of the Standard Model and its simple extensions. The
open issue, for any model, is whether these necessary conditions 
also amount to sufficient ones.
 
Although electroweak baryogenesis is by no means the only available
scenario for generating the BAU, it is attractive in that it offers for 
a very restricted framework, being successful only for specific
parameter values that can conceivably be probed at the LHC. A
disadvantage is that a complete and reliable calculation of the BAU 
is theoretically demanding (for a recent review see
ref.~\cite{mrm}). However, a crucial step, which by itself leads 
to strong constraints, is to prove the existence of a strong first-order 
electroweak phase transition. For this step, which is the topic
of the present paper, uncertainties can be brought under control 
through a combination of analytic computations and large-scale 
lattice Monte Carlo simulations.

Lattice studies of variants of the 
electroweak phase transitions have a long history by now. 
They represent the only known systematic way of circumventing 
the infrared problem of thermal field theory~\cite{linde} that 
limits the accuracy of perturbative evaluations.
For instance, it was originally
envisaged that  electroweak baryogenesis could work even within
the Standard Model~\cite{krs}, but nowadays it
is known that this possibility is not realized in nature. An
unambiguous reason is furnished by lattice 
simulations~\cite{endpoint,endpoint2}: unlike suggested 
by perturbation theory, the transition is a crossover for 
a Higgs mass compatible with either LEP or 
LHC~\cite{cms,atlas} bounds,
which implies that the system does not deviate from equilibrium.\footnote{%
  It is 
  commonly believed that the Standard Model also contains too little
  CP violation to allow for electroweak baryogenesis, however this 
  issue is harder to prove beyond doubt, cf.\
  e.g.~ref.~\cite{anders2} for recent work and references.}

On the other hand, simple extensions of the Standard Model,
particularly the Minimal Supersymmetric Standard Model (MSSM), might
change the picture.  Indeed, even though very strongly constrained by
now~\cite{cjm,cmp}, realizing electroweak baryogenesis in the classic
MSSM~\cite{Carena:1996wj}--\cite{mlo2} (or in extensions 
of the Standard Model resembling MSSM at low energies~\cite{dnq1}) 
seems not to be excluded~\cite{cnqw,clw,new}. In
particular, if right-handed stops are sufficiently light and
left-handed stops are heavy enough, the electroweak phase transition
can be strong even for a (lighter CP-even) Higgs mass $m_h\simeq
126$~GeV~\cite{cnqw, Carena:2008vj}. This assertion relies, however, 
on perturbation theory, whose accuracy cannot be taken for granted.
Therefore, to confirm or rule out the scenario, 
lattice Monte Carlo simulations appear welcome.

In the past, lattice analyses have extensively studied the electroweak
phase transition within the MSSM. However, even the latest
simulations~\cite{cpsim} only focussed on a part of the MSSM parameter space
that corresponds to a lightest Higgs mass $m_h \lsim 115$~GeV. This
seems to be excluded by the recent LHC data~\cite{Arbey:2012bp}.
It thus appears well-motivated to repeat the lattice analysis for 
$m_h \simeq 126$~GeV, and this is the aim of the present study.

The paper is organized as follows. In sec.~\ref{se:dim} we review the
dimensionally reduced effective theory of the MSSM with 
a light right-handed stop,
and choose a parameter region where perturbative calculations find a
strong electroweak phase transition. Within this region, we select a
parameter point with $m_h\simeq 126$ GeV where we analyze the phase
transition on the lattice (sec.~\ref{se:latt}). The lattice results
are compared with the perturbative ones in sec.~\ref{se:compare}. 
Finally, sec.~\ref{se:concl} is dedicated to an outlook and conclusions.

%
\section{Dimensionally reduced effective theory}
\la{se:dim}

For the benefit of an impatient reader, we start by summarizing the
four-dimensional (4d) parameter values which the effective theory
simulations are believed to correspond to (\se\ref{ss:params}).
Subsequently the theoretical foundations and practical implementation
of the dimensionally reduced (3d) effective theory construction are
briefly reviewed (\se\ref{ss:theory}).

%
\subsection{Parameter values}
\la{ss:params}

The present study is based on lattice simulations as explained in
ref.~\cite{cpsim}, and on analytic dimensional reduction and vacuum
renormalization formulae
as described in ref.~\cite{cpdr}. The physics setting
is different from what might appear ideal from today's perspective: 
in particular, the presence of a relatively light CP-odd Higgs mass 
$\mA \approx 150$~GeV, as well as of relatively light gluinos of mass
$M^{ }_3 \lsim 300$~GeV, were assumed in ref.~\cite{cpdr}. 
Even though this setting is problematic
because of light stop bounds (forbidding small
$M^{ }_3$~\cite{susybound}) and dark matter constraints 
(disfavouring light $\mA$~\cite{Kozaczuk:2011vr}),
it provides for a conservative framework to prove
the existence of a strong first-order phase transition
with $m_h \simeq 126$~GeV. Indeed, we expect to find a stronger
transition for larger CP-odd Higgs mass
(decreasing $\mA$ weakens the 
transition~\cite{cpsim, mAlarge}). Having relativistic
gluinos in the thermal bath increases the right-handed 
stop thermal mass by $\sim 20$\%
and consequently pushes the right-handed
stop vacuum mass parameter to more negative values but, 
as we have checked by a resummed 1-loop estimate, does otherwise
not significantly affect the phase diagram. 

A major difference with respect to ref.~\cite{cpdr} where a
left-handed stop mass $\mQ \lsim 1$~TeV was assumed, is that here we
push $\mQ$ to much larger values. This is needed to achieve
$m_h\simeq 126$~GeV but, as pointed out in ref.~\cite{lss}, this induces 
large logarithms that were not resummed in ref.~\cite{cpdr}. Our
evaluations of the Higgs and right-handed stop masses, $m_h$ and
$m_{\tilde t_R}$, as functions of the MSSM parameters, are therefore
approximate. 

Another important point is the renormalization scale at which 
the couplings appearing in the thermal mass corrections are evaluated.
Following ref.~\cite{hd}, it was assumed in ref.~\cite{cpdr} that 
the couplings run to a scale $\sim 2\pi T$ 
like in the Standard Model~\cite{generic}. 
To crosscheck the argument requires carrying out
a full 2-loop dimensional reduction computation, 
which is absent at present in the parameter range considered. 
The concrete consequence of the assumption
of a scale $2\pi T \gg m_\rmi{top}$ is that the strong gauge
coupling is quite small; this implies that the right-handed stop
squared mass parameter, $\mU^2$, does not need to be as
negative as sometimes assumed; and subsequently, that the physical
right-handed stop mass 
$ 
 m_{\tilde t_R} \simeq \sqrt{m_\rmi{top}^2 - \tmU^2}
$ 
(for $A_t \approx 0$ and $\tmU^2 \equiv - \mU^2>0$) is larger.

To be reminded of these uncertainties, we tag the parameters
mentioned, as well as the temperature, by a star in the following:
\be 
\quad \tmU^* \;, \quad \mQ^*
 \;, \quad m_h^* \;, \quad m_{\tilde t_R}^* \;, \quad T^* \;, 
\ee
and similarly for the less significant parameters. 
These numbers are therefore {\em not} to be interpreted 
as precise physical values.

With these reservations, the dimensional reduction formulae and the
notation are identical to refs.~\cite{cpsim, cpdr}. (As a small point,
explicit CP violation has been switched off for simplicity).  The
perturbative phase diagram is illustrated in fig.~\ref{fig:phase}.
For the parameter setting
\ba
 && 
 \tmU^* = 70.5~{\rm GeV} \;, \quad
  \mQ^*=7~{\rm TeV}  \;, \quad
  \mu^*=M_2^*=\mA^*=150~{\rm GeV} \;, \nn
 && \hspace*{2.5cm}
  \tan\beta^* = 15 \;, \quad A_t^*/\mQ^* = 0.02
 \;, 
 \label{setting}
\ea
where $\mu^*, M_2^*, \tan\beta^*$ and $A_t^*$ are defined in a standard 
way (see e.g.~ref.~\cite{cpdr}), 
the perturbative calculation yields $v(T_\rmi{c}^*)/T^*_\rmi{c} = 0.9$
in Landau gauge,
where $v(\Tc^*)$ is the gauge-fixed expectation value 
of the lighter Higgs at the critical
temperature $\Tc^*$ (the precise definitions of these observables are
given in secs.~\ref{se:latt} and \ref{se:compare}).\footnote{%
  The vev normalization is $v(0) \simeq 246$ GeV.}  
Notably, at this parameter point the lightest Higgs
pole mass is $m_h^* \simeq 126~$GeV 
within the 1-loop approximation~\cite{erz}.

Comparing with refs.~\cite{cnqw, Carena:2008vj}, the mass parameter
$\tmU^*$ and critical temperature $T^*_\rmi{c}$ evaluated at
$\mQ^* = 7$~TeV are substantially smaller, and consequently the
physical stop mass $m^*_{\tilde t_R}$ is larger. The difference
seems to be related to running effects in the thermal mass corrections 
as mentioned above (the discrepancy is smaller at $\mQ^* =
1$~TeV, but of course then $m_h^*$ becomes unphysically light).  It is
our ultimate goal to improve the dimensional reduction and vacuum
renormalization computations for the case of a very large $\mQ^*$, but
unfortunately this requires a substantial amount of new work, which is
postponed to future. If it turns out that our tagged parameters are
close to the physical ones, in particular 
$m_{\tilde t_R}\simeq 155~$GeV
(cf.\ \fig\ref{fig:phase}), then the tension between LHC data and 
the considered MSSM scenario would be 
relaxed~\cite{cjm, cmp, cnqw, susybound}.
(Note that since the transition we find is 
comfortably strong enough for baryogenesis, there remains some 
tolerance for a minor error in $m^{ }_{\tilde t_R}$.)

\begin{figure}[t]


\centerline{%
 \epsfysize=7.5cm\epsfbox{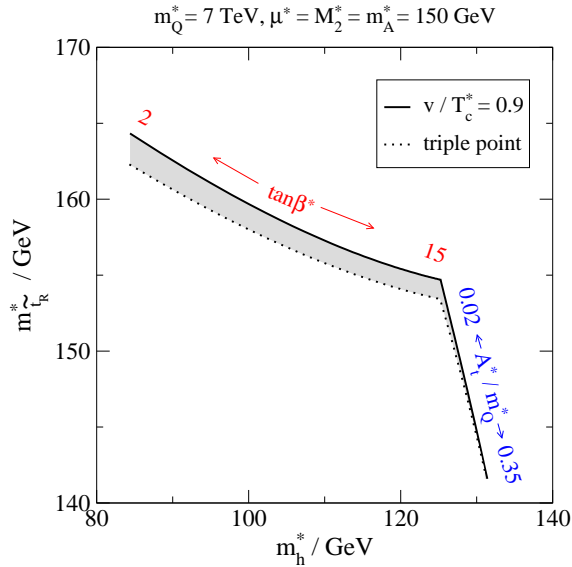}%
}

\caption[a]{\small Slices of the perturbative phase diagram of the
  effective theory considered, in terms of the MSSM-like parameters
  defined in the text. The upper boundary corresponds to parameters
  at which $v(\Tc^*) /\Tc^*$ equals 0.9 in the
  Landau gauge; the lower boundary to values below which the theory is
  driven to a colour-broken minimum according to the 2-loop effective
  potential.}

\la{fig:phase}
\end{figure}

Despite the above uncertainties, there is still a well-defined problem
to solve. Namely, we compare non-perturbative Monte Carlo simulations
with 2-loop perturbation theory {\em within the 3d effective
  theory}. In other words, the effective parameters entering the
simulations and the analytic formulae are identical.
Therefore the finding of a non-perturbative effect in one or the other
direction is likely to also persist with slightly modified parameter
values.\footnote{The conclusions of our analysis also apply to models
  beyond the Standard Model that resemble the considered effective
  theory at low energies, for instance certain
  triplet extensions of the MSSM \cite{dnq1} or the Standard Model
  with an extra coloured scalar~\cite{colorSM}.}
The actual simulations are carried out for the parameter values of
eq.~(\ref{setting}).

%
\subsection{Review of theoretical setting}
\la{ss:theory}

In order to appreciate the methods to be used, we briefly 
reiterate why it is non-trivial to determine reliably the properties
of phase transitions even in weakly coupled 
theories. Technically, this is due to the so-called infrared problem
of thermal field theory~\cite{linde}. Let $g$ denote a renormalized
gauge coupling. Then the theory has three different momentum scales: 
\begin{itemize}

\item
The scale $p\sim 2\pi T$, technically originating from fields
carrying non-zero Matsubara modes, is a ``hard scale''; perturbation
theory at this scale is free from infrared problems, and is expected
to converge well (practical tests in the Standard Model 
suggest that after a 2-loop computation the remaining errors 
are on the percent level~\cite{gR2}).

\item The scale $p\sim gT$, called the Debye or ``soft'' scale and
  technically originating from the interactions between the hard modes
  and the long-wavelength excitations, is also free from infrared
  problems. In general, perturbation theory at this scale converges
  slower than that at the hard scale, but in the electroweak theory
  there are very {\em many} hard modes, meaning that the Debye scale
  comes with a large numerical coefficient, making it in practice not
  much smaller than the scale $p\sim 2\pi T$.  Indeed it has been
  checked (by comparing with full 4d simulations in an SU(2)$+$Higgs
  toy model~\cite{buwu}) that even the soft scale can be integrated
  out with good precision in the Standard Model~\cite{gR2}.

\item
The ``ultrasoft'', or ``magnetic'' scale $p\sim g^2 T/\pi$, first
identified by Linde~\cite{linde}, is purely non-perturbative in 
nature, and needs to be studied numerically.

\end{itemize}
In the presence of a Higgs phase,
the $W^\pm,Z^0$ bosons have masses $\mW, \mZ \simeq gv(T)/2$.
If $v(T)\;\sim\; T$, than $\mW, \mZ$ are formally ``soft''
scales and perturbation theory may or may not work, depending on
numerical coefficients.  However, if $v(T) \; \lsim \; g T$, as is the
case on the side of the high-temperature  phase,
then perturbation theory certainly breaks down.  Therefore, in the
following, $\mW$ and $\mZ$ are treated as ultrasoft scales.

Now, even though a scale hierarchy of the type described is a problem
for perturbation theory (as it happens, it is also a challenge for direct
4d lattice simulations), it is a blessing once the problem is
rephrased in an effective field theory language. Indeed the hard and
soft scales {\em can} be integrated out perturbatively; it is only the
ultrasoft scale which needs to be studied with lattice simulations.
The integration out is called dimensional
reduction~\cite{dr1,dr2}, and the lattice simulations are then
those of purely bosonic 3d gauge$+$Higgs theories.

It is worth stressing that even if {\em no} simulations were carried
out, it would nevertheless be useful to organize the computation in the
above language. The reason is that the integrations over the hard and
soft scales implement all-orders resummations (such as the ``daisy''
one), which are in fact necessary even for observables not sensitive
to the ultrasoft scale.

With this background, the general steps of the adopted approach are as
follows:
\begin{itemize} 

\item
{\bf Derivation of a 3d effective theory.} The first step is 
to integrate out the hard and soft scales.   
To be specific, this step involves not only the actual 
finite-temperature calculations but also the corresponding 
vacuum computations, in order to express the $\msbar$ scheme parameters 
in terms of experimentally measurable quantities. 

\item
{\bf 2-loop perturbation theory within 3d effective theory.}
Once an effective theory is available, it can first be studied
with 2-loop perturbation theory. This is the level that previous 
experience from the Standard Model and MSSM has shown to be semi-quantitatively
accurate, {\em provided} that the transition is strong enough. 
Analytic expressions also yield a qualitative 
understanding of various parametric dependences. 

\item
{\bf Lattice formulation of 3d effective theory.}
The derivation of the effective theory and its 2-loop perturbative analysis
are, as a rule, carried out in a continuum regularization scheme, 
such as the $\msbar$ scheme. Obviously, a lattice provides for 
a regularization scheme of its own. For a systematic study,
the two schemes need to be related to each other; the principal
relations have been worked out for a large class of theories~\cite{akr}.  

\item
{\bf Numerical simulations within 3d effective theory.}
The last step of the program is to carry out lattice simulations. 
Although these are substantially simpler than full 4d simulations, 
they do remain non-trivial: infinite-volume and continuum limits 
need to be carefully taken in order to obtain physical results. 

\end{itemize}

Without going into further details, which have been explained in
refs.~\cite{cpsim,mssmsim}, we recall the continuum form of the
3d theory simulated. The theory contains two Higgs SU(2) doublets,
$H_1$ and $H_2$, and a field $U$ which is SU(2) singlet but SU(3)
triplet. The Lagrangian has then the most general form allowed by
symmetries,
\ba
 {\mathcal{L}^\rmii{ }_\rmii{3d}}\, {T^*}  & \equiv & 
 \fr12 \tr G_{ij}^2 + (D_i^s U)^\dagger (D_i^s U) + 
 m_\rmii{$U$}^2(T^*)\, U^\dagger U
   + \lambda^{ }_\rmii{$U$} (U^\dagger U )^2 \nn
 & + & 
 \gamma^{ }_1\, U^\dagger U \, H_1^\dagger H^{ }_1 + 
 \gamma^{ }_2\, U^\dagger U \, \tilde H_2^\dagger \tilde H^{ }_2
 + \bigl[\gamma^{ }_{12}\, U^\dagger U \,
    H_1^\dagger \tilde H^{ }_2+\Hc\bigr] \nn
& + & \fr12 \tr F_{ij}^2 
 + (D_i^w H^{ }_1)^\dagger(D_i^w H^{ }_1) + 
   (D_i^w H^{ }_2)^\dagger(D_i^w H^{ }_2) \nn
& + &  m_1^2(T^*)\, H_1^\dagger H^{ }_1
 +  m_2^2(T^*)\, \tilde H_2^\dagger \tilde H^{ }_2 
+ \bigl[ m_{12}^2(T^*)\, H_1^\dagger \tilde H^{ }_2 +\Hc  \bigr]  
 \nonumber \\[1.5mm]
& + &  \lambda^{ }_1 (H_1^\dagger H^{ }_1)^2 + 
 \lambda^{ }_2 (\tilde H_2^\dagger \tilde H^{ }_2)^2 + 
 \lambda^{ }_3 H_1^\dagger H^{ }_1 \tilde H_2^\dagger \tilde H^{ }_2 + 
 \lambda^{ }_4 H_1^\dagger \tilde H^{ }_2 \tilde H_2^\dagger H^{ }_1 
 \nonumber \\[1.5mm]
& + & \bigl[ \lambda^{ }_5 (H_1^\dagger \tilde H^{ }_2)^2 + 
 \lambda^{ }_6 H_1^\dagger H^{ }_1 H_1^\dagger \tilde H^{ }_2 +
 \lambda^{ }_7 \tilde H_2^\dagger \tilde H^{ }_2 
               H_1^\dagger \tilde H^{ }_2 + \Hc\bigr]
 \;, \la{MSSM_action} 
\ea
where a factor $T^*$ has been inserted in order to keep
4d dimensionalities of fields and couplings; 
$D_i^s$, $D_i^w$ are the SU(3) and SU(2) covariant derivatives;
$G^{ }_{ij}$, $F^{ }_{ij}$ the corresponding field strength tensors;
and $ \tilde H^{ }_2 = i \sigma^{ }_2 H_2^* $.  The hypercharge
coupling has been set to zero, so there is only a global U(1) symmetry. 
The gauge couplings are denoted by
$g_w^2$ and $g_s^2$ for SU(2) and SU(3),
respectively. The actual values 
corresponding to the setting of eq.~\eqref{setting} are
\ba
& & 
 m_1^2(T^*) \approx 26504 \mbox{ GeV}^2 + 0.1311 (T^*)^2 \;,
 \la{mm1} \\
 & & 
 m_2^2(T^*) \approx -4004 \mbox{ GeV}^2 + 0.6311 (T^*)^2 \;,
 \la{m22} \\
 & & m_{12}^2(T^*) \approx -1481 \mbox{ GeV}^2 
 - 0.0133 (T^*) ^2 \;, \la{m12} \\
 & & 
 \mU^2(T^*) \approx -4958 \mbox{ GeV}^2 + 0.8607 (T^*)^2 \;,
 \la{mmU}  \\
 & & 
 \gamma^{ }_1 \approx -0.0031\;, \quad
 \gamma^{ }_2 \approx  0.9995\;, \quad
 \gamma^{ }_{12} \approx -0.0018\;, \quad
 \lambda^{ }_\rmii{$U$} \approx  0.2020\;, \\
 & & 
 \lambda^{ }_1 \approx  0.0649\;, \quad
 \lambda^{ }_2 \approx  0.1491\;, \quad
 \lambda^{ }_3 \approx  0.0589\;, \quad
 \lambda^{ }_4 \approx -0.1784\;, \\
& & 
 \lambda^{ }_5 \approx  0.00009\;, \quad
 \lambda^{ }_6 \approx -0.00113\;, \quad
 \lambda^{ }_7 \approx -0.00117
 \;, \\
& & 
 g_w^2 \approx 0.418 \;, \quad
 g_s^2 \approx 1.085
 \;. \la{3dprms}
\ea
As can be seen from \eqs\nr{m22}, \nr{mmU}, it is the fields
$\tilde H^{ }_2$ and $U$ that ``drive'' the transition; correspondingly, 
the couplings $\gamma^{ }_2$, $\lambda^{ }_\rmii{$U$}$, 
and $\lambda^{ }_2$ are the most significant ones. 

%
\section{Lattice simulations}
\la{se:latt} 

%
\subsection{Action and algorithms}

The theory in \eq(\ref{MSSM_action}) is discretized in
the standard way, and for implementation details, we refer to
refs.~\cite{cpsim,mssmsim}.  For the gauge action we use the usual single
plaquette Wilson formulation, with SU(2) and SU(3) lattice couplings
$\beta_w = 4/(g_w^2 T^* a)$ and $\beta_s= 6/(g_s^2 T^* a)$.  Here $a$ is the
lattice spacing, which we parameterize through $\beta_w$ from now on.

Only the bare mass terms require renormalization 
in a 3d super-renormalizable theory:
\be 
 m^2_\rmi{latt} = m^2  + \Delta m^2\;.  
\ee
Here $m^2$ stands for either $\mU^2,m^2_1,m^2_2$ or $m^2_{12}$
in the $\msbar$ scheme, and $\Delta m^2$ is the counterterm
containing the linear and logarithmic in $a$ divergences~\cite{cpsim}. 
With this renormalization only $O(a)$ cutoff errors remain, 
and the continuum limit can be taken by performing simulations at 
different $\beta_w$ and by extrapolating $\beta_w \rightarrow\infty$
afterwards.

The update algorithm is a combination of heat bath and overrelaxation
updates, with one compound update sweep consisting of one heat bath
and three overrelaxation sweeps.  The measurements are performed and
recorded every two compound update steps.

\begin{table}
\centerline{%
\begin{tabular}{cl}
\hline
$\beta_w$ & volumes \\
\hline
~8        & $12^3$, $16^3$ \\
10        & $16^3$  \\
12        & $16^3$, $20^3$, $32^3$, $12^2\times 36$, $20^2\times 40$ \\
14        & $24^3$, $14^2\times 42$, $24^2\times 48$ \\
16        & $24^3$, $16^2\times 48$, $20^2\times60$, $24^2\times 72$ \\
20        & $32^3$, $20^2\times 60$, $26^2\times72$, $32^2\times64$ \\
24        & $24^3$, $32^3$, $48^3$, $24^2\times78$, $30^2\times72$ \\
30        & $48^3$ \\
\hline
\end{tabular}}
\caption[a]{\small
  The simulation volumes used in multicanonical simulations at temperatures
  around $\Tc^*$.}
\la{table:volumes}
\end{table}

It turns out that at the critical temperature the transition is
strongly of first order, and the system does not spontaneously tunnel
from one metastable phase to another with standard simulation
algorithms.  However, the probability distribution at the tunnelling
region needs to be accurately resolved in order to precisely determine
the critical temperature, the order parameter discontinuities, and the
surface tension.  We handle the strong metastability through the use
of the {\em multicanonical} simulation method with automatic weight
function calculation, combined with reweighting of the simulation
temperature.  We use the average $H_2^\dagger H^{ }_2$ as the
multicanonical order parameter since this condensate is the most
sensitive to the phase transition ($H_1$ is practically inert due to
the large value $\tan\beta^*=15$). The volumes simulated are listed in
table~\ref{table:volumes}.

%
\subsection{Observables and results}

The observables measured on the lattice are all extracted from
gauge-invariant operators and are therefore gauge-independent by
construction. Here we reiterate their definitions and show the 
main results.

\paragraph{Condensates as functions of the temperature:}

\begin{figure}[t]


\centerline{%
 \epsfysize=6.5cm\epsfbox{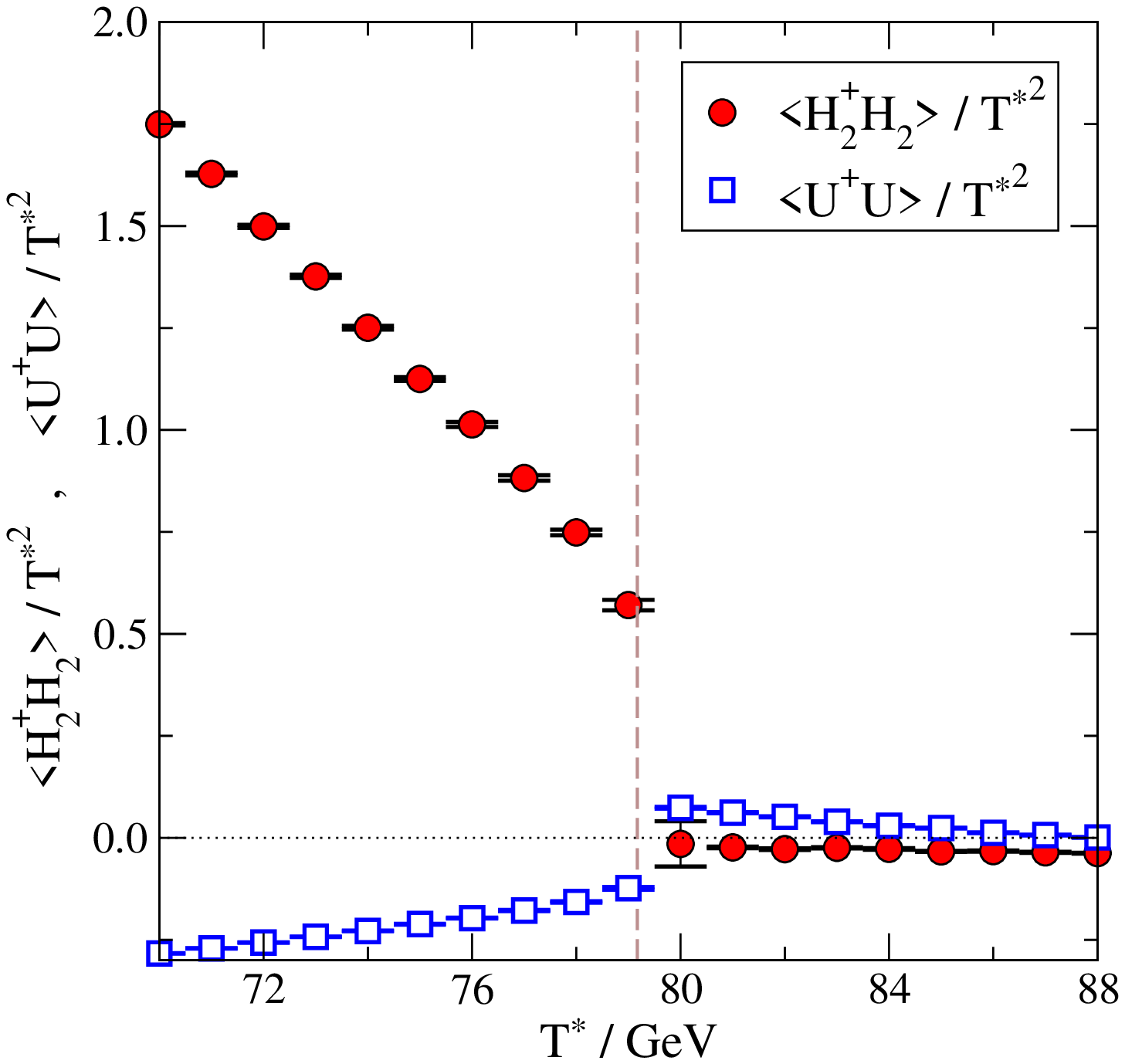}%
 \qquad \epsfysize=6.5cm\epsfbox{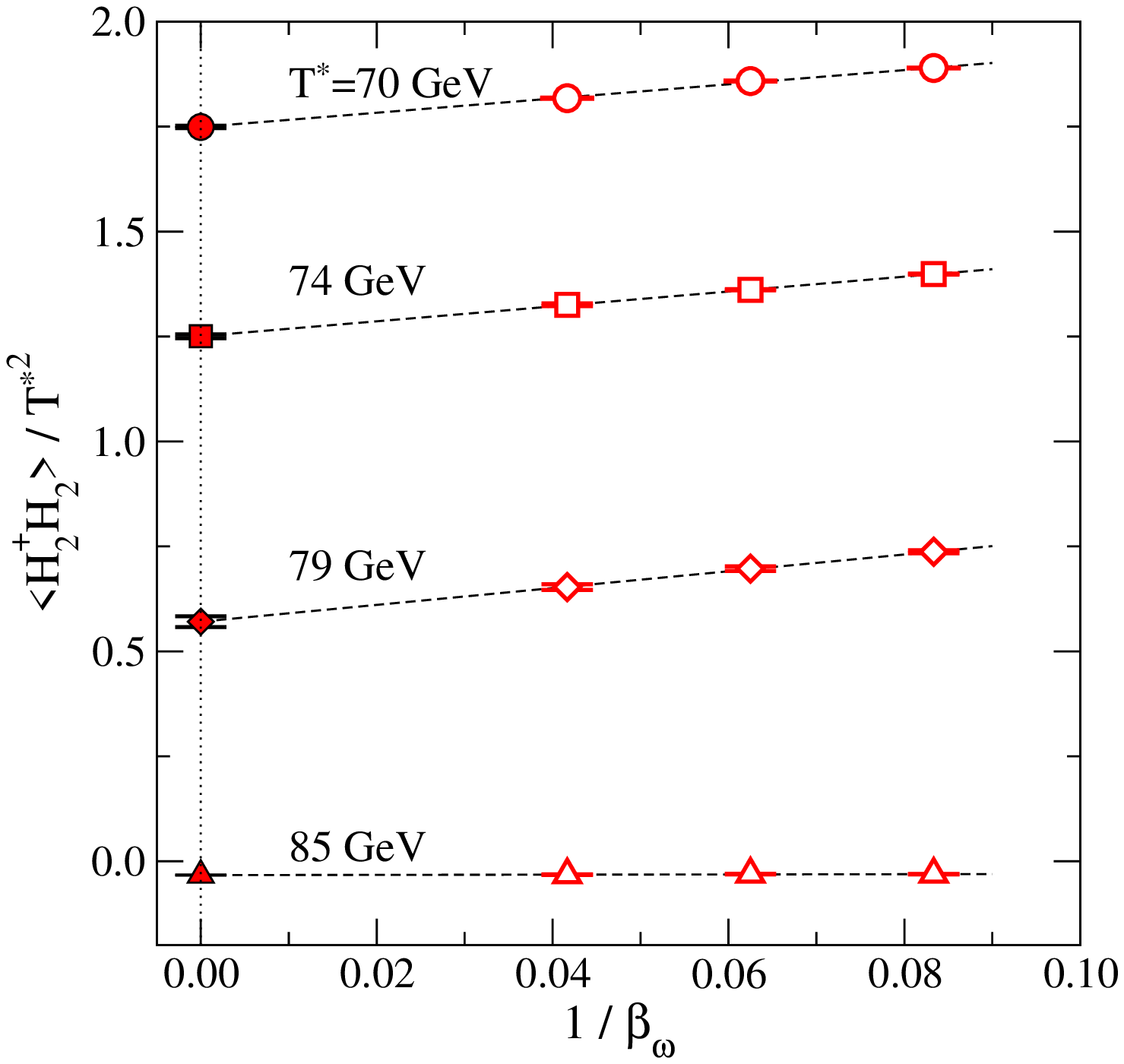}
}

\caption[a]{\small Left panel: the expectation values of $\langle
  H_2^\dagger H^{ }_2\rangle$ and $\langle U^\dagger U\rangle$ as
  functions of the temperature $T^*$ in the continuum limit.  The
  vertical dashed line shows the phase transition temperature.  Right
  panel: continuum extrapolation of $\langle H_2^\dagger H^{ }_2\rangle$
  at four chosen temperatures.}
\la{fig:tscan}
\end{figure}

In order to obtain an overall view of the behaviour of the condensates
$\langle H_i^\dagger H_i^{ }\rangle$ and $\langle U^\dagger U\rangle$
we perform a series of simulations at 
$T^* = 70$--$90$ GeV at three different lattice spacings, $\beta_w = 12$,
$16$ and $24$.  The values of these condensates in 
the $\msbar$ scheme with the scale parameter $\bmu = T^*$ are obtained
by subtracting lattice divergences, for instance 
\be \frac{\langle
  H_2^\dagger H^{ }_2\rangle}{(T^*)^2} =  
 \frac{\langle H_2^\dagger H^{ }_2\rangle_{\rm latt}}
 {(T^*)^2} - \frac{\Sigma}{2\pi a T^*} -
  \frac{3 g_w^2}{16\pi^2} 
  \left[ \log\frac{6}{aT^*} + 0.66796\right] +
 O(a) \la{condsub} \;,
\ee
where $\Sigma = 3.1759...\,$ originates from a 3d lattice 
tadpole integral.
Here and in the following, we refer to the condensates 
with their original 4d dimensionalities.
For our parameter settings, the values of these condensates
extrapolated to the continuum are shown in the left panel of
fig.~\ref{fig:tscan}. The figure highlights a first-order phase
transition at a temperature around 80 GeV 
(the transition temperature as determined with 
separate multicanonical simulations is indicated with 
the vertical dashed line; see discussion below).
The stop field $U$ is also affected by the transition, 
due to its strong coupling to $H_2$.

On the right panel we show the continuum extrapolation at four
selected temperatures.  The lattice divergences have been subtracted
according to \eq\nr{condsub}, whereas the remnant $\rmO(a)$ effects 
have been eliminated by a linear extrapolation in $1/\beta_w$.


\paragraph{Critical temperature ($\Tc^*$):}

\begin{figure}[t]


\centerline{%
 \epsfysize=7.0cm\epsfbox{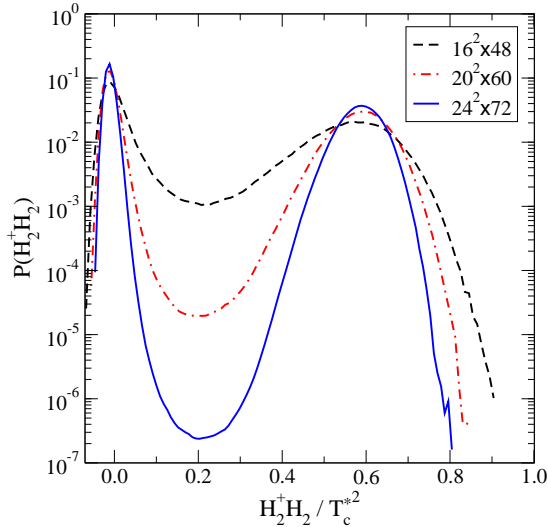}%
}
\caption[a]{\small The volume-averaged probability distribution of
  $H_2^\dagger H^{ }_2$ at three different volumes
  with $\beta_w=16$ and at $T_{{\rmi c},\beta_w=16}^*$. 
  As the volume increases, the probability density between 
  the two peaks decreases exponentially.}
\la{fig:hg}
\end{figure}

The critical temperature is defined by the value of $T^*$ at which 
two phases, identified through the expectation value of 
$\langle H_2^\dagger H^{ }_2 \rangle $, are equally likely to exist.  
Because the
tunneling between the two phases is strongly suppressed (it is exponentially
suppressed at large volumes, cf.\ \eq\nr{sigma_def}), 
multicanonical simulations are implemented to
overcome the tunneling barrier.

In our case the critical temperature is expected to be at around 80
GeV, as already indicated by fig.~\ref{fig:tscan}. Near
this temperature we therefore run multicanonical simulations for the
volumes and lattice spacings listed in table~\ref{table:volumes}.  For
most of the lattice spacings, several volumes are used, enabling us to
crosscheck the absence of finite-volume effects.  
The number of measurements at each
volume is around ($0.5$ -- $2)\times 10^6$.  Cylindrical volumes are
needed for the surface tension measurement, as described below.

\begin{figure}[t]


\centerline{%
 \epsfysize=7.0cm\epsfbox{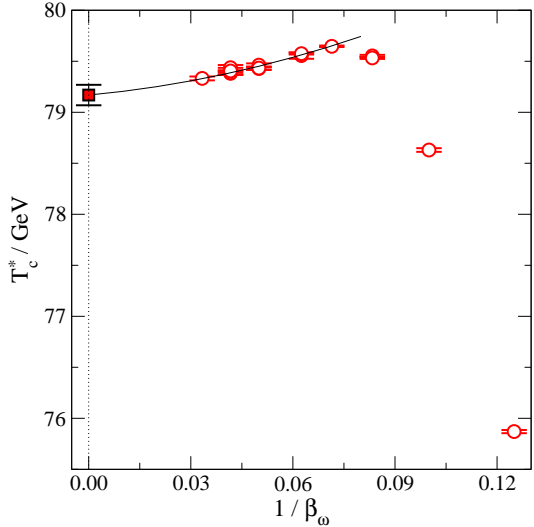}%
}
\caption[a]{\small The critical temperature as a function of the
  lattice spacing $1/\beta_w$. With the volumes shown in table
  \ref{table:volumes} no systematic volume dependence is seen, and all
  volumes are included in the plot.  The interpolating curve is a
  second-order polynomial in $a$, fit to the points $\beta_w \ge 14$.}
  \la{fig:tc}
\end{figure}

As an example, the distribution of the volume-average
of $H_2^\dagger H^{ }_2$, 
resulting from the simulations performed at $\beta_w=16$,
is shown in fig.~\ref{fig:hg}.
Clearly, the probability of configurations between the 
two phase peaks becomes strongly suppressed as the volume
increases.  The temperature has been reweighted from the simulation
temperature ($T^*=79.5$\,GeV) to the apparent critical temperature where
the area of the peaks is equal. The critical temperature averaged over
all three volumes is $T_{{\rmi c},\beta_w=16}^*\simeq 79.57$\,GeV.

All the critical temperatures obtained for the lattice spacings and
volumes of table~\ref{table:volumes} are summarized in
fig.~\ref{fig:tc}. A clear dependence on the lattice spacing is
present, but results tend to stabilize as the continuum limit is
approached. We then fit only data at $\beta_w \ge 14$ by a
second-order polynomial in $1/\beta_w$. The continuum 
intercept reads
\be
 \Tc^* = 79.17 \pm 0.10\,\mbox{GeV}\;, \la{tc} 
\ee
with $\chi^2/\mbox{d.o.f.} = 11.2/8$ for the fit.

\begin{figure}[t]


\centerline{%
 \epsfysize=7.0cm\epsfbox{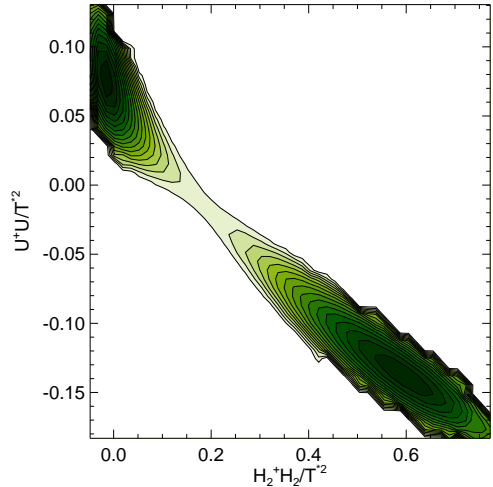}%
}
\caption[a]{\small
  Contour plot of the joint probability distribution of
  the volume-averaged $H_2^\dagger H^{ }_2$ and $U^\dagger U$ at 
  $\Tc^*$.  The measurement is for $\beta_w=20$,
  volume $=32^2\times 64$. The $\msbar$ scheme
  value of $\langle U^\dagger U \rangle$ is slightly negative
  in the low-temperature phase (cf.\ the text). 
}
\la{fig:hg2}
\end{figure}

In fig.~\ref{fig:hg2} the joint probability distribution of the 
volume-averaged $H_2^\dagger H^{ }_2$ and $U^\dagger U$ is shown.
Clearly, the condensates move together: when $H_2^\dagger H^{ }_2$ 
becomes large, this effectively increases the mass of the $U$-field 
through the interaction mediated by the coupling $\gamma^{ }_2$
in \eq\nr{MSSM_action}, and therefore its fluctuations become smaller. 
In fact in the $\msbar$ scheme with the scale choice $\bmu = T^*$
the value of $\langle U^\dagger U \rangle$ becomes slightly 
negative at low temperatures, but this simply means 
that $U$ is tightly confined into its
``symmetric'' phase in this regime.\footnote{%
  It is in principle possible that separate symmetric
  $\leftrightarrow$ broken $H_2$ and $U$ phase transitions exist,
  giving rise to three phases: fully symmetric, broken $H_2$, and
  broken $U$.  This has indeed been observed in a different parameter
  region, see fig.~9 of ref.~\cite{cpsim}.  However, in the case 
  at hand we do not observe a stable broken $U$ phase
  (no scan in $\tmU^*$ has been performed here).}


\paragraph{Higgs discontinuity ($v(\Tc^*)$):} 

The Higgs discontinuity is defined on the lattice in a gauge-invariant and
scale-independent manner as
\be
 \biggl( \frac{v^2(\Tc^*)}{2} \biggr)^{ }_\rmi{latt} \equiv
 \Delta \Bigl\langle \sum_{i=1}^2 
 H_i^\dagger H^{ }_i
 \Bigr\rangle
 \;. \la{vev_def} 
\ee
The quantity
$
 \Delta \langle \cdot \rangle \equiv 
 \langle \cdot \rangle_\rmi{broken} - 
 \langle \cdot \rangle_\rmi{symmetric}
$ 
is measured from the probability distributions at the critical
temperature (such as those shown in fig.~\ref{fig:hg}) by integrating
over the peaks independently.\footnote{%
  This involves setting a ``separatrix'' between the two phases, 
  which is chosen to lie at the minimum of the probability distribution. 
  As can be deduced from \fig\ref{fig:hg}, the ambiguity related to the 
  choice becomes exponentially insignificant at large volumes.}  
In our case, the contribution of $H_1$ can be
neglected since $\tan\beta^*\gg 1$.

The Higgs discontinuity as a function
of the lattice spacing is shown in fig.~\ref{fig:delta} (left
panel). Each point is obtained by averaging over the volumes listed in
table~\ref{table:volumes}.  In this case $v(\Tc^*)/\Tc^*$ levels off
almost completely at small enough lattice spacing, and a linear fit to
points with $\beta_w \ge 16$ gives
\be 
 \frac{v(\Tc^*)}{\Tc^*} = 1.117 \pm 0.005\;, 
\ee
with $\chi^2/\mbox{d.o.f.}  = 2.8/3$.

\begin{figure}[t]


\centerline{%
  \epsfysize=6.5cm\epsfbox{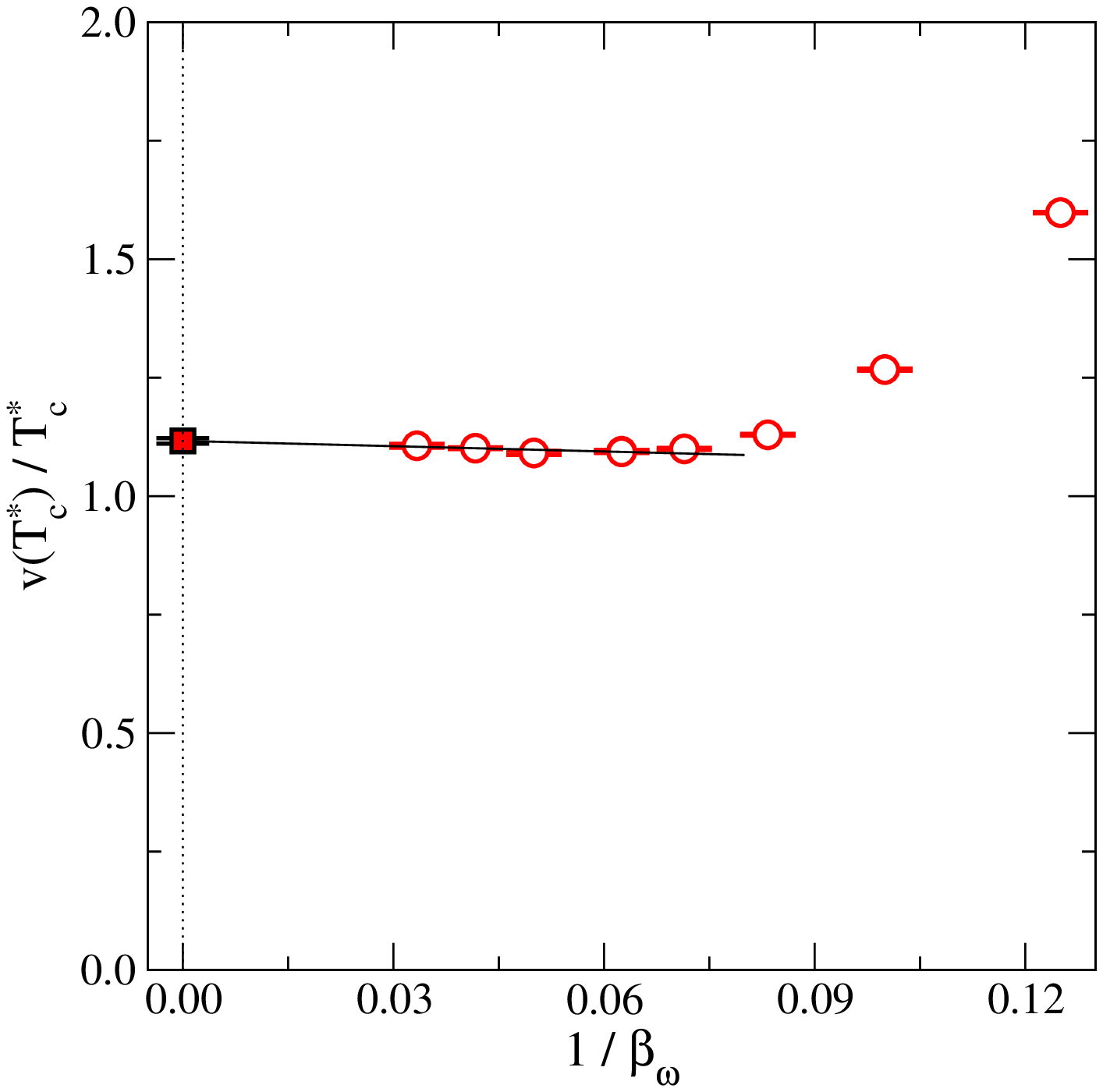}%
 \qquad  \epsfysize=6.5cm\epsfbox{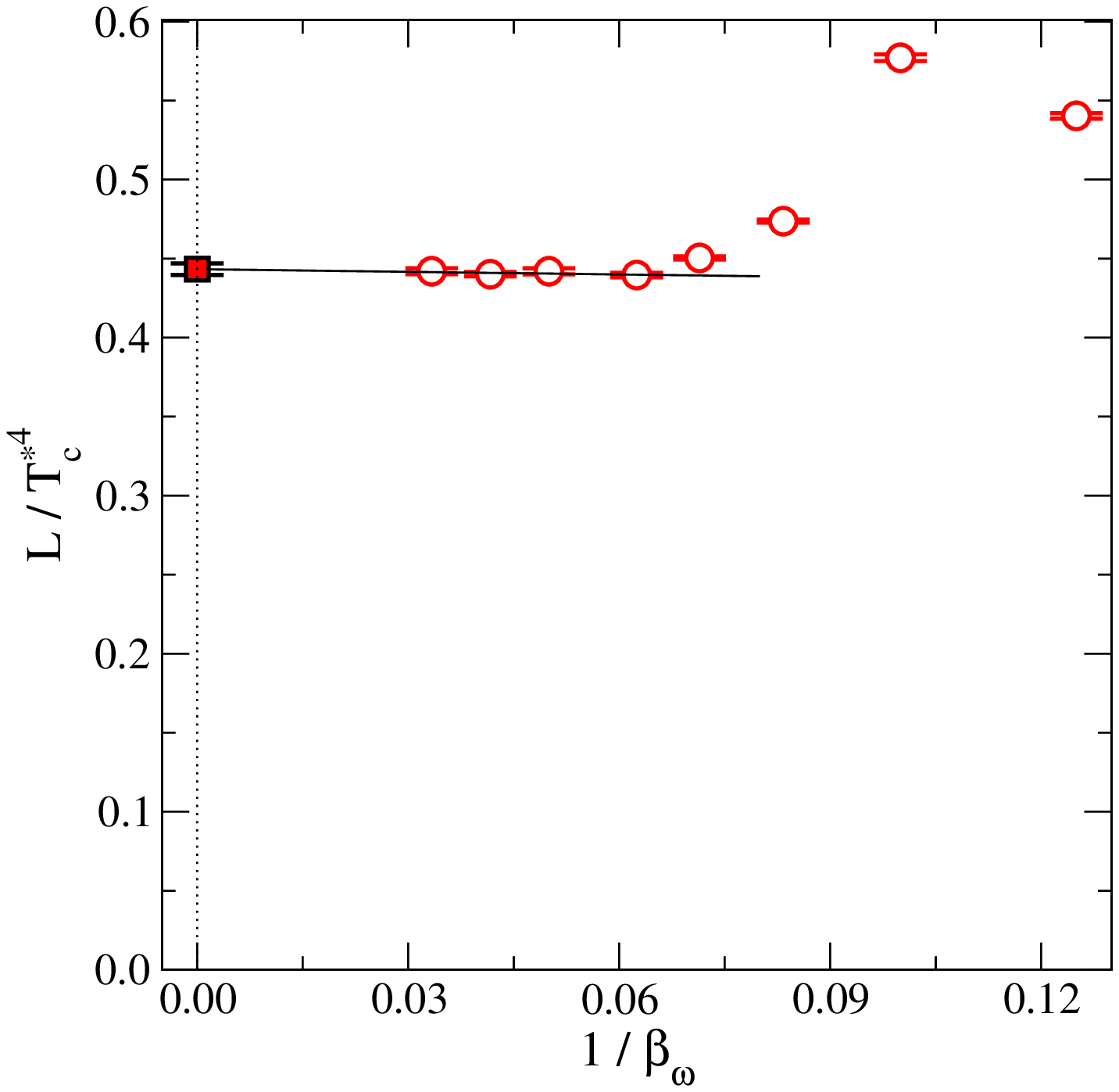}%
}
\caption[a]{\small The Higgs discontinuity (left panel) and the latent
  heat (right panel) as functions of the lattice spacing.  The linear
  continuum fit to points at $\beta_w \ge 16$ (solid line) is also
  shown.}
  \la{fig:delta}
\end{figure}

The value of $v(\Tc^*)/\Tc^*$ is large enough so that
the sphaleron rate (which is proportional to 
the baryon number violation rate) after 
the phase transition is negligible.
Nevertheless, it should be stressed that 
$v(T^*)$ as defined by \eq\nr{vev_def}
is a purely thermodynamic quantity, and
therefore is not {\em equivalent} in any strict sense to the
sphaleron rate. However, as non-perturbative real-time simulations for
the Standard Model have shown~\cite{anders}, the value of $v(\Tc^*)/\Tc^*$
is {\em strongly correlated} with the sphaleron rate.  Given the fact
that non-perturbative real-time simulations are very expensive, only
$v(\Tc^*)/\Tc^*$ is measured in the present study.


\paragraph{Latent heat ($L$):} 

The latent heat is defined as the discontinuity of the energy density
across a first order phase transition. It plays an essential role in the
real-time hydrodynamics of bubble nucleation and growth. Within the
approximate parametrization of the 3d theory used in this work, in
which only mass parameters depend ``non-conformally'' on the
temperature [cf.\ \eqs\nr{mm1}--\nr{3dprms}], it can be measured as
\ba
 \frac{ L }{(\Tc^*)^3}  & = &     
 \Delta \biggl\langle
  { U^\dagger U } 
  \,  
  \fr{{\rm d}}{{\rm d} T^*}
 \biggl[  \fr{m_\rmii{$U$}^2(T^*)}{(T^*)^2} \biggr]  +
 \sum_{i=1}^2 
 { H_i^\dagger H^{ }_i } 
 \,  
 \fr{{\rm d}}{{\rm d} T^*} 
 \biggl[ \fr{m_i^2(T^*)}{(T^*)^2} \biggr]
 \nn & & \quad + \,    
  \bigg(
 { H_1^\dagger \tilde H^{ }_2 } 
 \, 
 \fr{{\rm d}}{{\rm d} T^*}
 \biggl[ \fr{m_{12}^2(T^*)}{(T^*)^2} \biggr]
  + \mbox{H.c.} \bigg)
 \biggr\rangle 
 \la{L_def}
 \;.
\ea
The discontinuity in \eq\nr{L_def} is readily
measured from multicanonical simulations, with the results shown on
the right panel of fig.~\ref{fig:delta}.  The continuum limit gives
\be \frac{ L }{ (\Tc^*)^4 } = 0.443 \pm 0.004 \;, \ee
with $\chi^2/\mbox{d.o.f} = 1.01/2$ using a linear fit to data with
$\beta_w \ge 16$. 


\paragraph{Surface tension ($\sigma$):} 

\begin{figure}[t]


\centerline{%
  \epsfysize=7.0cm\epsfbox{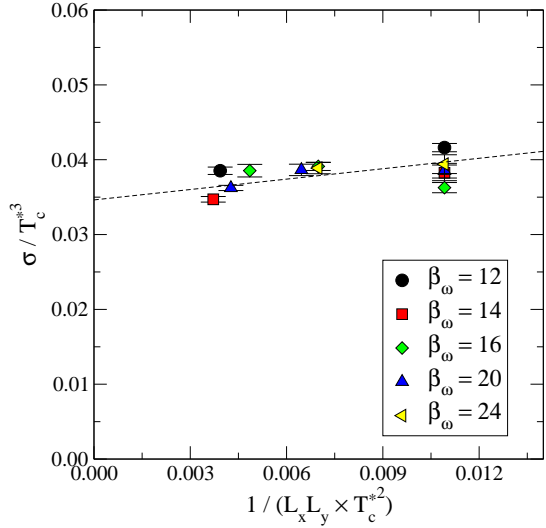}%
}
\caption[a]{\small The surface tension of the interface between the
  symmetric and broken Higgs phases, plotted versus the inverse area
  of the interface.}
  \la{fig:sigma}
\end{figure}

The surface tension is defined as the additional free energy per area
carried by an interface between the two co-existing phases. This means
that, in the large-volume limit, the probability of a configuration
which contains an interface of area $A$, denoted by $P_\rmi{min}$, is
smaller than the probability of a configuration without interfaces,
$P_\rmi{max}$, by a factor 
\be 
 \frac{P_\rmii{min}}{P_\rmii{max}} = 
 \exp \Bigl( {-\frac{\sigma A}{ T^*}} \Bigr) 
\;. \la{sigma_def} 
\ee 
When the volume is finite, there are corrections to this relation, and
properly accounting for these accelerates the convergence to the
infinite-volume limit (cf.\ refs.~\cite{cpsim,mssmsim} for details).

The probabilities
${P_\rmii{min}}$ and ${P_\rmii{max}}$ can be directly read from the
distributions in fig.~\ref{fig:hg}:
${P_\rmii{max}}$ is the average peak height and ${P_\rmii{min}}$ is
the minimum between the peaks.  Because of the periodic boundary
conditions the configurations here contain (at least) two interfaces.
In practice it is advantageous to use lattice volumes where one
dimension is longer than the other two, e.g. cylindrical volumes
($L^{ }_z \gg L^{ }_x,L^{ }_y$).  In
this case the interfaces are oriented transversely to  
the long direction ($A \approx 2 L^{ }_x L^{ }_y$). 

In fig.~\ref{fig:sigma} measurements of the surface
tension at each of the cylindrical volumes and lattice spacings
indicated in table~\ref{table:volumes} are shown.  
In this case we are not able
to obtain independently 
reliable (i.e.\ with small $\chi^2$/d.o.f) infinite volume
and continuum limit extrapolations.  This is likely due to still
remaining finite volume effects; surface tension measurements are
notoriously sensitive to volume.  However, it is clear from the plot
that the measurements settle down on a narrow band, independently of
the lattice spacing.  Extrapolating the band to the infinite volume we
cite a conservative but unprecise estimate of the error:
\be
 \frac{\sigma}{(\Tc^*)^3} = 0.035 \pm 0.005\;.  
\ee
%

%
\section{Comparison with perturbation theory}
\la{se:compare}

With a view of learning about generic features of the dynamics of the
theory, probably applicable also to other parameter values than the
very ones considered here but nevertheless close to $m_h\simeq 126$
GeV, we proceed to comparing the lattice results with those of 2-loop
perturbation theory within the 3d theory. We stress that since both
results are based on the {\it same} 3d theory, the comparison is {\it
  not} jeopardized by perturbative uncertainties in dimensional 
reduction and vacuum renormalization as discussed in 
\se\ref{ss:params}. Indeed, these ultraviolet features play 
a role only in the relation of the approximate parameters 
($T^*, m_h^*, m_{\tilde t_R}^*$, etc) to the physical ones 
($T, m_h, m_{\tilde t_R}$, etc). 
For conceptual clarity, we furthermore
split the comparison into two parts, given that 
some of the perturbative numbers cited are 
specific to Landau gauge, in accordance with established
(although not necessary) conventions of the field.

\begin{figure}[t]


\centerline{%
\epsfysize=7.5cm\epsfbox{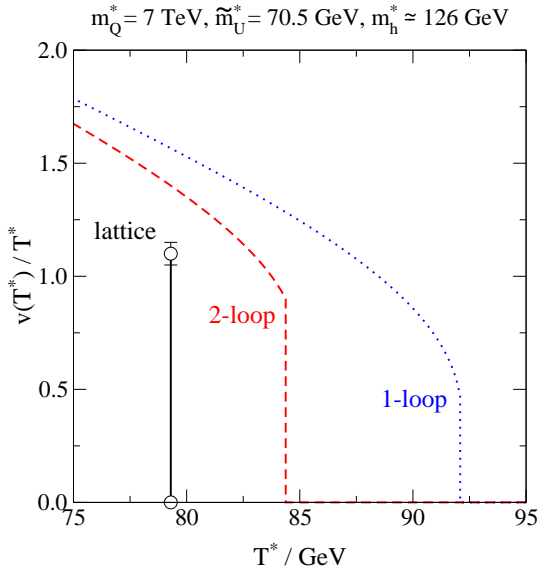}%
}

\caption[a]{\small Comparison of perturbative and lattice results for
  the properties of the phase transition (here $v(T^*)$~refers,
  strictly speaking, to different quantities on the two sides; cf.\
  \se\ref{ss:corr}).  }

\la{fig:broken}
\end{figure}

%
\subsection{Identical observables}
\la{ss:strict}

Two of the observables, namely the critical temperature and latent heat, have
definitions [see \eq\nr{L_def} for the latter] that can be operatively applied
both to lattice and perturbative calculations. Their values for 
the case analyzed are
\ba 
\frac{\Tc^*}{\mbox{GeV}} \; = \; 79.17(10) \quad \mbox{(latt)} \;, 
& & \frac{\Tc^*}{\mbox{GeV}} \; = \; 84.4 \quad \mbox{(pert)} \;,
\la{Tc_comp} \\[3mm] 
\frac{L}{(\Tc^*)^4} \; = \; 0.443(4) \quad \mbox{(latt)} \;, 
& & \frac{L}{(\Tc^*)^4} \; = \; 0.26 \quad \mbox{(pert)} \;. 
\la{L_comp} 
\ea
For the critical temperature the situation is illustrated in
\fig\ref{fig:broken}.  As has been observed also in the past~\cite{cpsim}, the
main qualitative effect from non-perturbative dynamics is that the critical
temperature is lowered. The latent heat is enhanced by$~\sim 50$\%.

%
\subsection{Correlated observables}
\la{ss:corr}

Within 2-loop perturbation theory, the gauge-independent observable 
defined by \eq\nr{vev_def} happens to be very close to the gauge-fixed 
Higgs vev as computed in Landau gauge. Due to the fact that 
the Landau-gauge convention continues to be widespread in the 
literature, we therefore compare the lattice number directly
with the Landau-gauge perturbative result: 
\be
 \biggl( \frac{v}{\Tc^*} 
 \biggr)^{ }_\rmi{\eq\nr{vev_def}} 
 \; = \; 1.117(5)
 \quad \mbox{(latt)} 
 \;, \qquad
 \biggl( \frac{v}{\Tc^*} \biggr)^{ }_\rmi{Landau} \; = \; 0.9
 \quad \mbox{(pert)}
 \;. \la{vev_comp}
\ee
The percentual strengthening effect is smaller than for $L$, 
because $L$ is essentially quadratic in $v$ (cf.\ \eq\nr{L_def}).
Another quantity for which we are influenced by convention and 
ease of computation is the surface tension; this is 
usually extracted from the Landau gauge effective potential, with 
tree-level kinetic terms employed in finding the saddle point 
solution (for a recent discussion, see ref.~\cite{Garny:2012cg}). 
The comparison reads
\be
 \biggl( \frac{\sigma}{(\Tc^*)^3} \biggr)^{ }_\rmi{\eq\nr{sigma_def}}
 \; = \; 0.035(5)
 \quad \mbox{(latt)} 
 \;, \qquad
 \biggl( \frac{\sigma}{(\Tc^*)^3} \biggr)^{ }_\rmi{Landau} 
 \; = \; 0.025
 \quad \mbox{(pert)}
 \;. \la{sigma_comp}
\ee
%

%
\section{Discussion and conclusions}
\la{se:concl}

The recent LHC discovery of a Higgs-like boson with a mass of around
126 GeV may have provided crucial information for electroweak
baryogenesis. In many models beyond the Standard Model, the success of
electroweak baryogenesis in explaining the baryon asymmetry of the
Universe is indeed very sensitive to the Higgs mass through the
requirement of a strong first-order electroweak phase transition.  In
the MSSM, perturbative studies have suggested that a strong
first-order electroweak phase transition may exist even at $m_h\simeq
126$ GeV~\cite{cnqw,Carena:2008vj}. However, since the transition is
not exceedingly strong and the side of the ``symmetric'' phase is
purely non-perturbative in nature, it is not clear whether the
perturbative predictions are quantitatively accurate.  In this paper
we have studied the infrared dynamics of the transition by simulating
a dimensionally reduced effective theory numerically, and compared the
results with 2-loop perturbative calculations within the same
effective theory.  Unless there are unexpectedly large 2-loop
corrections to the relations between the effective parameters of the
dimensionally reduced theory and four-dimensional physical low-energy
observables, the simulations correspond to an MSSM-like parameter
point with $m_h\simeq 126$ GeV and $m^{ }_{\tilde t_R} \simeq
155$~GeV, with larger uncertainties on the latter. (The
  simulations, however, are expected to also cover other extensions of
  the Standard Model having a stop-like field at the electroweak
  scale, i.e.\ providing for a similar low-energy effective theory.)

In the lattice simulations carried out, we have consistently seen a
stronger transition than in perturbation theory. Actually, despite the
larger Higgs mass, the strengthening effect is more substantial than
in ref.~\cite{cpsim}. In some sense the system
is driven towards a phase where the right-handed 
stop experiences very strong fluctuations
(manifested by a large $\langle U^\dagger U \rangle$), 
and the transition to the electroweak minimum
takes place ``from there'' (cf.\ \fig\ref{fig:hg2}).\footnote{%
  Considerations such as those in ref.~\cite{cm} are
  evaded because this is not a perturbatively ``colour-broken''
  phase and the transition to the physical vacuum
  does take place, as the estimate in \eq\nr{Tnucl} shows.}

For a precise understanding of baryogenesis, it is not enough to study 
the properties of the transition at the critical temperature, but issues
such as supercooling, nucleation, and bubble dynamics need to be
considered as well (see e.g.\ refs.~\cite{no}--\cite{ay} 
for recent discussions).
The nucleation temperature, $T^{ }_{\rm n}$, is well approximated by the 
classical estimate~\cite{landau5} if it is calculated with non-perturbative 
values of the latent heat and 
surface tension inserted~\cite{gdm}. After nucleation
and bubble collisions, the latent heat released may also reheat the system
towards the critical temperature, which would enhance baryon number washout. 

Various scenarios for the real-time dynamics of the transition have been 
studied in ref.~\cite{ksl}, and in fact the dynamics of the present 
transition is not unlike case~(A) considered there.
More precisely, reheating up to
$\Tc$ would take place if
\be \frac{L}{\Tc^4} \; \gsim \; 8 \biggl( \frac{\sigma}{\Tc^3}
\biggr)^{3/4} \;.
  \ee
This is (narrowly) avoided according to \eqs\nr{L_comp} and
\nr{sigma_comp}.  Moreover, supercooling is roughly
\be \frac{\Tc^{ } - T^{ }_\rmi{n}}{\Tc^{ }} \; \simeq \; 0.54 \,
 \biggl( \frac{\sigma}{\Tc^3} \biggr)^{3 / 2} \biggl( \frac{\Tc^4}{L} \biggr)
 \simeq 0.008 \;. 
\label{Tnucl}
\ee
It is therefore quite modest, and is not expected to change the Higgs
vev substantially (cf.\ \figs\ref{fig:tscan}, \ref{fig:broken}). 
However, modern
hydrodynamics studies of the phase transition will make these
conclusions firmer~\cite{hydro}.

At the moment the physical 4d parameter values to which our
simulations correspond, particularly the right-handed stop mass,
contain uncertainties of several GeV. To remove this perturbative
uncertainty, full 2-loop dimensional reduction and 
at least 1-loop on-shell
vacuum renormalization computations (expressing $\msbar$ scheme
parameters in terms of physical low-energy observables), such as were
carried out for the Standard Model~\cite{generic}, are needed.

Despite the encouraging results that we have found, it is also clear
that an exclusion of light SU(2)-singlet stop-like particles
at the LHC could easily
rule out the MSSM-based electroweak baryogenesis scenario.  Therefore,
it may be worthwhile to apply the techniques of the present study to
more general models.  Very many possibilities can be envisaged; as an
example of a relatively well-constrained one, let us mention the
so-called Inert Doublet Model~\cite{id1,id2}, i.e.\ a particular
version of the two-Higgs-Doublet Model with an imposed unbroken Z(2)
symmetry which reduces the number of free parameters.  This model has
many attractive features, for instance the heavy Higgs doublet could
naturally serve as Dark Matter~\cite{mhgt1,mhgt2}. The theory could
conceivably lead to a strong first order phase transition as
well~\cite{senj,jc1,gch}, and a non-perturbative study may again be
welcome.

%
\section*{Acknowledgements}

K.R.~acknowledges support from the Academy of Finland project 1134018.
Simulations were carried out at the University of Helsinki
and at the Finnish IT Center for Science (CSC).


\end{document}